\documentclass[manuscript]{aastex}

\shorttitle{Galaxy Formation and Evolution}
\shortauthors{Burstein \& Blumenthal}

\begin{document}

\title{The Role of a Hot Gas Environment on the Evolution of Galaxies}

\author{David Burstein\altaffilmark{1},
George Blumenthal\altaffilmark{2}}

\altaffiltext{1}{Department of Physics and Astronomy, Box 871504, Arizona
    State University, Tempe, AZ  85287-1504}
\altaffiltext{2}{Board of Studies, Astronomy \& Astrophysics,
University of California, Santa Cruz, CA 96054}

\begin{abstract}


Most spiral galaxies are found in galaxy groups with low velocity
dispersions; most E/S0 galaxies are found in galaxy groups with
relatively high velocity dispersions. The mass of the hot gas we can
observe in the E/S0 groups via their thermal X-ray emission is, on
average, as much as the baryonic mass of the galaxies in these groups.
By comparison, galaxy clusters have as much or more hot gas than
stellar mass.  Hot gas in S-rich groups, however, is of low enough
temperature for its X-ray emission to suffer heavy absorption due to
Galactic HI and related observational effects, and hence is hard to
detect. We postulate that such lower temperature hot gas does exist in
low velocity dispersion, S-rich groups, and explore the consequences of
this assumption. For a wide range of metallicity and density, hot gas
in S-rich groups can cool in far less than a Hubble time.  If such gas
exists and can cool, especially when interacting with HI in existing
galaxies, then it can help link together a number of disparate
observations, both Galactic and extragalactic, that are otherwise
difficult to understand.

\end{abstract}


\section{Introduction}

Both the Nearby Galaxies Catalog \citep{tul87} and the discovery of
X-ray emitting hot gas in galaxy groups \citep[][hereafter
JSM00]{mdmb93,hp00,jsm00} have provided strong evidence that most galaxies
reside in galaxy groups, not unlike our own Local Group (e.g., all of
$\sim 2500$ galaxies in the Nearby Galaxies Catalog are within 1 Mpc
of a galaxy group or cluster).  Indeed, the evidence suggests that up
to 90\% of galaxies reside in galaxy groups, rather than in the
better--studied galaxy clusters \citep[cf.][]{tul87,rn93}.

Galaxy groups can be divided into two types by the morphology of their
giant galaxies \citep{mdmb96a,dbetal97}.  The dividing line chosen by
\citet{mdmb96b} is that any group that has more than 1/3 of its giant
galaxies of types gE or S0 is called E/S0-rich; all others are called
S-rich.  Two-thirds of the \citet{rn93} nearby sample of galaxy groups
are S-rich by this definition; 1/3 are E/S0-rich \citep{dbetal97}.
The mean observed velocity dispersion in the galaxy groups classified
by Nolthenius decreases by a factor of two from E/S0-rich to S-rich
groups \citep{mdmb96b}.  Subsequently, \citet{zm98} and \citet{mz98}
found that while E/S0-rich groups can contain 30 or more galaxies more
luminous than $\rm M_B \sim -16$, typical S-rich groups, such as our
own Local Group, contain only 4-8 galaxies this luminous.

\citet{mdmb96b} point out that the higher velocity dispersion of
E/S0-rich groups corresponds to a hot gas temperature of $\sim 0.9$
keV, which is easily seen with ROSAT.  In contrast, for S-rich groups,
the predicted temperature based on their lower velocity dispersion
averages 1/4 of this value, making any X-ray emission from hot gas
subject to severe extinction by metals associated with HI gas in our
own Galaxy \citep{rs77}.  

Here we make the reasonable assumption that low velocity dispersion,
S-rich galaxy groups {\it do} contain hot gas (in some S-rich groups
such hot gas has been found; cf. JSM00; Xu \& Wu, 2000). In Section 2
we show that, for a range of temperature, metallicity and density,
much of the hot gas in galaxy groups can cool in far less than a
Hubble time. In Section 3, we briefly discuss the consequences of the
interrelationship that must exist among the dark matter distribution
and hot gas in galaxy groups, and the consequences that relationship
holds for the formation and evolution of galaxies in the groups.
Section 4 discusses some of the varied observations of galaxies and
galaxy groups that the assumptions of this Letter can help explain or
predict.

\section{Cooling Times for Hot Gas in Galaxy Groups}

The measured temperatures of X-ray emitting gas in galaxy groups range
from $T \sim 0.3 - 1.0$ keV, with a strong correlation between the
velocity dispersion of the groups and their hot gas temperatures
(e.g., JSM00).  This lower temperature limit arises not from the
absence of groups with low velocity dispersions (predominantly
S-rich), but rather from the He and metal nuclei absorption
(associated with Galactic HI) of X-ray emission from hot gas with $T
\le 0.3$ keV ($T \le 3.5 \times 10^6$ K) \citep[e.g.,][]{rs77}.  The
fact that lower hot gas temperature means lower luminosity, and that
the X-ray background is higher for $\rm T < 0.5$ keV, also contribute
to the lack of X-ray detected low velocity dispersion groups.

\begin{figure}
\epsscale{0.8}
\plotone{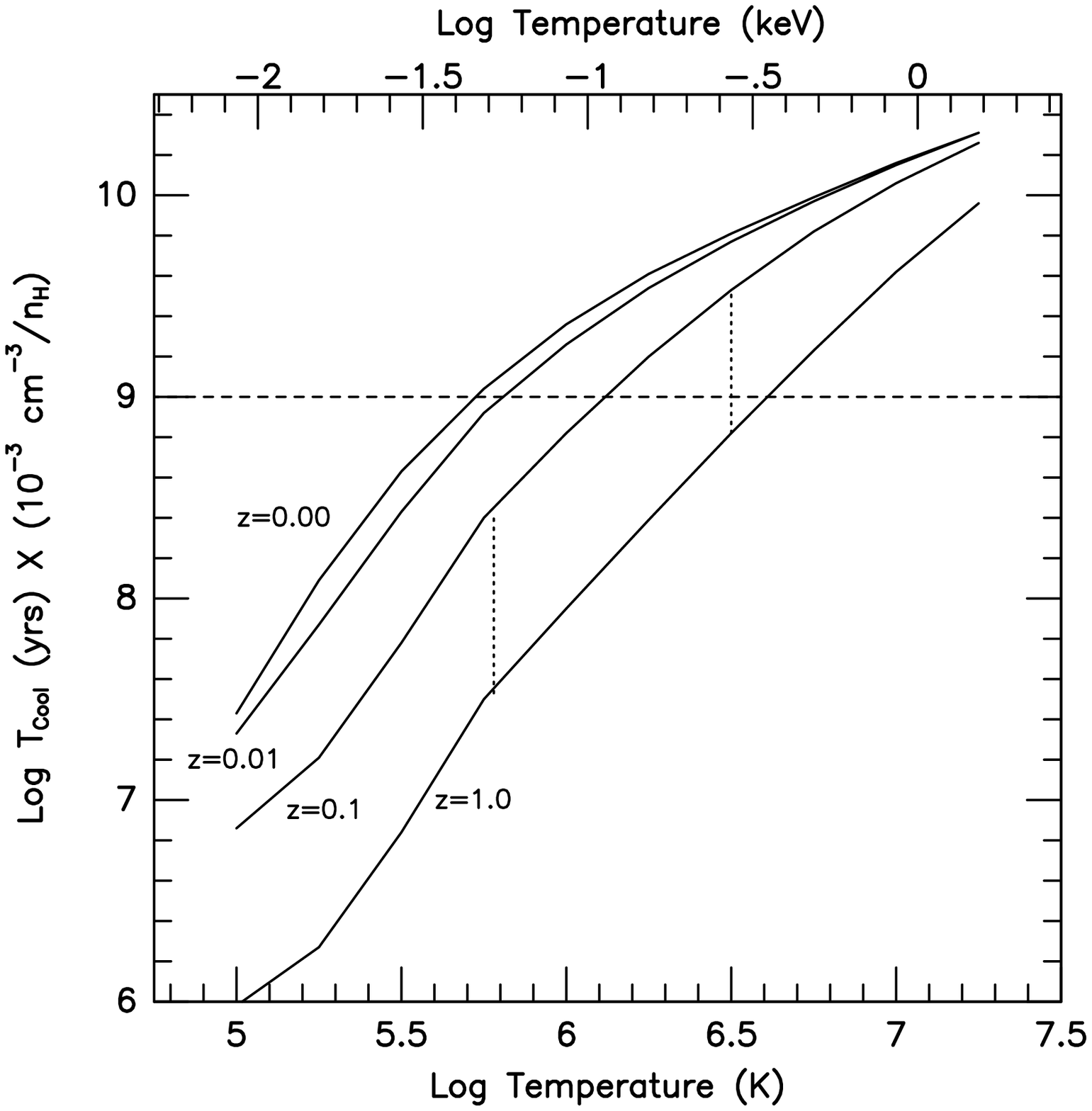}
\figcaption{The logarithm of the cooling rates (in years) of hot gas
as a function of the logarithm of its temperature (given both in terms
of K and in keV), for four values of metallicity relative to solar
($z$); 1.0, 0.1, 0.01 and 0.00. Dotted lines are drawn between the z =
1.0 and z = 0.1 curves at values of Log T = 5.78 and 6.5,
corresponding to the range of observed velocity dispersions in
low-$\sigma$ galaxy groups, for a gas density of $\rm n_H = 10^{-3} \,
cm^{-3}$.  A dotted line at a cooling time of 1 Gyr is drawn to guide
the eye. \label{fig1}}
\end{figure}

It is well-known that X-ray emitting gas with $\rm T \sim 10^6 - 10^7$
K can cool efficiently under the ``right'' circumstances of gas
density, metallicity and temperature \citep{nsf84}.  In
Figure~\ref{fig1} we plot the gas cooling times as a function of
temperature for a range of temperature and metallicity relevant for
gas in galaxy groups, using the results given by \citet{sd93}. Cooling
times are inversely proportional to the density of gas; the density
($\rm n_H = 10^{-3} \, cm^{-3}$) used for Figure~\ref{fig1}
corresponds to the typical gas densities seen in the interiors of the
E/S0-rich galaxy groups \citep[cf.][]{mdmb96a}.  JSM00 finds that most
galaxy groups have hot gas with metallicities $z > 0.1$.  Even for gas
densities a factor of ten less than used for Figure~\ref{fig1},
cooling times are still less than a Hubble time for $z > 0.1$,
especially for the gas temperatures ($\rm T < 3.5 \times 10^6$ K) we
expect to find in S-rich groups.

\section{Hot Gas and Galaxy Evolution}

The existing data is consistent with the mass of hot X-ray emitting
gas being comparable to the visible mass of galaxies, stars and gas
(cf. discussion in JSM00). The fact that hot intragroup/intracluster
gas contains metals implies that any primordial gas that was
originally there was subsequently enriched from star formation.
Consequently, during the history of these galaxies, their star
formation processes must have produced a comparable amount of mass in
hot gas released into the intragroup medium as there is mass in the 
stars themselves today.  The corresponding production rate for hot gas
in stellar evolution implies an average star formation rate not wildly
different from what we see in our Galaxy today \citep[cf.][]{bach96}.

By creating this hot gas via stellar evolution, galaxies are putting a
substantial amount of their mass into ``hot storage'', where
subsequent cooling and interactions with galaxies may significantly
affect the galaxies' structure and evolution.  This is yet another reason
\citep[cf. the density-morphology relation for galaxies,][]{ad80} to
view the formation and evolution of galaxies not in individual terms,
but rather in the context of the galaxy groups and clusters to which
they belong.

The mass-to-light ratios for galaxy groups, derived from the X-ray
observations, are similar to those for galaxy clusters, 100-120 in
solar units, B mag (cf. JSM00), implying dark matter domination.  The
mean velocity dispersion for S-rich groups is 1/2 that of E/S0-rich
groups (cf. above), and E/S0-rich groups also have effective radii
that are 50\% larger than S-rich groups \citep[cf. data in][]{dbetal97}.
On average, E/S0-rich groups are more massive (6$\times$) and denser
(1.7$\times$) than S-rich groups.  Because of its tight correlation
with velocity dispersion, the hot gas temperature in galaxy groups is
determined by the density and mass of the group, which in turn is
dominated by dark matter.

It is also noteworthy that the velocity dispersions {\it within} large
galaxies in groups tend to be comparable to the velocity dispersions
of the galaxies {\it in} their groups: E/S0 galaxies have typical
velocity dispersions (at $\rm L^*$) of 250 km s$^{-1}$, while giant
spirals typically have equivalent velocity dispersions of 150-200 km
s$^{-1}$ \citep{dbetal97}. Obviously, this trend breaks down at the
higher masses of clusters with their much higher velocity dispersion,
but there is evidence that clusters represent the end point of the
clustering of groups in dense environments \citep{bfpr84}.  Within
galaxy groups, the ``dark matter/visible matter conspiracy''
corresponding to the similarity of stellar motions and dark matter
velocity dispersion therefore applies to both the environment of
galaxies and to the galaxies themselves \citep[cf.][]{br85}.

This view is consistent with the hierarchical-clustering-merging (HCM)
picture \citep{dbetal97} in which E/S0 galaxies are formed in
denser, more massive environments, while spirals are formed in lower
mass, less dense environments. The new wrinkle here is something that
the HCM scenario has trouble predicting quantitatively (owing to a
lack of physical understanding of the origin of the initial mass
function for stars). Namely, galaxies put much of their baryonic
mass into ``hot'' storage.  Some of this hot gas will be in storage in
the halo of the galaxies themselves, while some will be within the
general group medium.

The fact that much of this hot gas can cool efficiently to cold gas in
less than a Hubble time can have important observational implications.
The interaction of hot intragroup gas with the cold gas in galaxies
moving within the groups can lead to complex hydrodynamical effects
including shocks and phase transitions, with subsequent cooling. This
interaction of hot and cold gas can lead to several interesting
galactic phenomena (e.g., tidal tails, galactic fountains, small,
HI-rich dwarf galaxies in orbit about their parent galaxies, such as
in the Local Group).  In fact, shocks that form in the interaction
region lead to higher densities for the hot gas, with consequently
shorter cooling times.  Quantitatively addressing these complications
will involve detailed modeling with yet-to-be-written N-body hydro
codes, but it is reasonable to assume that the cooling of this hot gas
and its interactions with extant galaxies can significantly affect the
evolution of these galaxies. We now explore some of the plausible
observational consequences of this interaction.

\section{Observational Consequences of Galaxy Interactions with
Intragroup Gas}

For the following, we make three assumptions: a) hot gas exists in all
galaxy groups; b) much of this gas can cool in much less than a Hubble
time; and c) the cooling rate is even faster when there are shocks
associated with the interstellar medium in galaxies.  These
assumptions lead to a wide range of linked observational implications.

1. The density/morphology relation for galaxies: It has long been
known \citep[e.g.][]{ad80} that galaxy morphology correlates with
local initial conditions such as mass density. Almost certainly, part
of this correlation is causal, since higher density implies more major
merging, which probably leads to ellipticals. At the same time, higher
initial densities lead to groups with higher velocity dispersion and
gas temperature. While the gas in S-rich groups will therefore be able
to cool in less than a Hubble time, the hotter gas E/S0-rich groups
may be too hot ($\sim 10^7$ K, cf. JSM00 and Fig. 1) to efficiently cool.

2. Galaxy warps, tidal tails and HI mass surface density
distributions: Most edge-on spiral disks have warps that persist well
into their extended HI distribution \citep{sanc76}, which is
consistent with the warp model developed to match the HI distribution
\citep{bosma81}. Surprisingly, there is evidence that the ratio of
mass surface density (mostly dark matter) to HI surface density in the
outer regions of spirals is a constant over several scale lengths
\citep{bosma81,hsv01}, with a sharp outer cut-off in the HI surface
density (Hoekstra et al.). Some HI-dominated tidal tails extending
from spiral galaxies have more gas than likely just from the
galaxies from which they are extended \citep{hy01}.

These observations of spiral galaxies are consistent with a strong
continuing interplay between cold HI in galactic disks and a cooling,
hot intragroup medium. In particular, hot gas may cool to form a part of
the HI disk, or there may be a pressure equilibrium between the colder
disk material and the hot gas. Tidally-extruded HI from galaxies can
possibly stimulate formation of HI from the hot gas that surrounds the
galaxies, likely forming the HI in pressure equilibrium with the hot
gas.  Only detailed hydro modeling can tell us more about how this
interaction might work, but the prima facie evidence is that there
is simply too much mass in HI tidal tails to have come just from the
outer, low HI surface density parts of spirals. The observed sharp 
cutoffs in the HI disks of galaxies might also arise from dynamical 
pressure or cooling from the hot gas surrounding the galaxy.

3. High velocity HI clouds: These exist near our Galaxy, with
their distances and origin still hotly debated \citep{blitzetal97,
sem02a,sem02b}. Recent observations \citep[cf.][]{sem02a,sem02b}
indicate that the high velocity clouds near our Galaxy are enveloped
in hot gas of temperature $\sim 10^6$ K.  Similar kinds of ``high
velocity clouds'' are observed near the disks of other spiral galaxies
\citep{sancetal01}, suggesting that this phenomenon is not isolated.
It is possible that these cold clouds could arise from thermal cooling
stimulated by the passage of dwarf galaxies or by an outflowing wind
driven by star formation in galaxy disks.  Similarly, when we look at
the Galactic poles in both directions, we see HI gas infalling toward
the Galactic plane \citep[cf.][]{crt71}.  This is consistent with the
idea that a local blowout for a galaxy disk can stimulate cold gas to
form from the hot gas in the galaxy's halo, and fall back to the 
Galactic plane.

4. Ultraviolet spectroscopic observations of high ionization
absorption features in distant objects show the presence of hot gas
near our Galaxy along different lines of sight
\citep[e.g.,][]{ddf95,sswrm02,sem02b} as well as in other galaxy
groups \citep[e.g.,][]{tripp02}. As pointed out by \citet{mdmb96b},
this is precisely what hot gas in S-rich groups (or in our galactic
halo and/or in our Local Group) would produce.

5. Galaxy groups which show strong tidal interactions among their
members often have a substantial fraction of their HI outside their
galaxies.  For example, the M81 group has 25\% of its HI gas in its
intragroup medium, rather than in the galaxies \citep{ads81}.  This
does not include the HI gas from the intragroup medium that has
already been dumped on M82 and NGC~3077 and formed stars there.  It is
unlikely that tidal tails alone can produce such a high percentage of
HI gas in the general group medium (again, they arise from the
relatively low density outer parts of galaxies).  Much of the
intragroup gas in the M81 group could have come from cooling of the
hot gas in that group, stimulated by the tidal interactions of the
galaxies, as well as with subsequent star formation in M82 and NGC 3077.

6. The solution of the ``G-dwarf problem'' in our Galaxy most likely
requires a constant infall of low metallicity gas \citep[cf.][]{cmg97}.
Similarly, semi-analytic models of galaxy evolution seem to imply the
need for continuous infall of gas from outside galaxies to produce the
galaxies we see today \citep[cf.][]{sem02a}.  Hot gas in the halo or
the immediate environment of spiral galaxies, which is of moderately
low metallicity, and which can cool to cold gas, can provide just such
a reservoir.

7. Spiral arms appear in HI disks of spirals outside the disks of
stars \citep[e.g.,][]{bosma81}.  Given that galaxies move at roughly
the speed of sound though their intragroup gas, the shocks that are
produced can help stimulate spiral arm formation in the HI gas in the
outer parts of galaxies.

8. The relationship between kinematic velocity fields and HI
distributions in spiral galaxies is complicated.  While the kinematic
velocity fields in many non-interacting spiral galaxies are typically
very regular, the angular distribution of HI gas (and, to some extent,
the disk stars) can be very irregular \citep[e.g.,][]{bosma81}. In
extreme cases, the HI density distribution within a spiral galaxy disk
can be very lopsided, while preserving a regular velocity field. There
are exceptional cases in which the kinematic velocity fields can also
be lopsided \citep{sssv99}. If cold gas is being produced by cooling
of hotter gas due to interactions with, for example, gas produced by
star formation in the galaxy disks, this could easily lead to a highly
variable gas density without significantly affecting the kinematic
velocity field.  This also suggests that the disks form from the
inside out (higher gas density in the galaxy interiors), producing
blue color gradients owing both to lower metallicity and younger stars
as a function of increasing radius \citep[cf.][]{ch95}.

Of course, the above is not an exhaustive list of observational
consequences of our assumptions. Moreover, not every one of these
observations is necessarily a consequence of our assumptions.
Rather, we suggest that all of the observations listed here are {\it
consistent} with our assumptions.  Based on this discussion, we have
estions for future investigation:

i. Density influences the formation and evolution of galaxies in more
than just defining the circumstances of their birth.  Rather, the
temperature of the hot gas they produce in their groups also can
dictate their subsequent evolution. Hydro modeling of galaxies
must take into account the environments in which we find them, which
are group-dominated.

ii. Observations indicate a range in gas mass in galaxy groups
(JSM00), some of which may be observationally-driven, but some of
which may be real. Separately, spiral galaxies are found with a wide
range in surface brightness \citep[cf.][]{impeyetal96}.  There could
be a relationship between gas mass/gas temperature ranges in galaxy
groups and the surface brightness of their spiral galaxies that should
be explored.

iii. Similarly, HI content and properties of spiral galaxies would
depend on the galaxy groups in which we find them.  The known large
variation of HI properties among spiral galaxies \citep{bosma81,
hsv01} makes this prediction only testable with a large statistical
sample of galaxies in a wide range of group environments.
Unfortunately, most of the nearby, large spirals which have been
well-studied for their HI distribution are part of the Coma-Sculptor
Cloud, which is completely comprised of S-rich groups \citep{tul87}.

iv. Of course, direct detection of hot gas in low dispersion galaxy
groups would directly prove our assumptions. Unfortunately, the very
long exposure times necessary for this to happen make such
observations hard to do.

If the way galaxy formation and evolution proceeds as we have outlined
in this Letter is correct, then two key pieces of information are
missing from high-z galaxy images: the three-dimensional distribution
of dark matter and the kinds of hot gas that are found at each stage
of galaxy formation and evolution.  Knowing the temperature and
density of hot gas in groups as a function of redshift would allow one
to make better observational predictions for different ranges of time.

In closing, it is clear that many of the possible observational
consequences we have identified can be modeled numerically. In
principle, N-body hydro codes can adequately model the interactions of
hot gas in galactic groups with the disks of spiral galaxies. However,
the star formation process and the way that star formation relates to
galaxy (and surrounding group) properties may turn out to be an
essential element. Until we understand the physics that produces the
initial mass functions of stars, any hypotheses such as ours that
involve interaction of hot gas in galaxy groups with cold gas and star
formation in galaxies must remain at least partly speculative.

\acknowledgements

We wish to thank Dr. George Coyne, SJ, Dr. Jose' Funes, SJ, Dr. Enrico
Corsini and Professor Francesco Bertola for organizing the June 2000
meeting on ``Galaxy Disks and Disks in Galaxies'', and conversations
with Renzo Sancisi, which greatly stimulated our ideas.

\clearpage

\end{document}